\documentclass{aipproc}
\def\selectedlayoutstyle{8x11single}
\layoutstyle\selectedlayoutstyle

\def\kpc {\rm{Kpc}}

\def\beq{\begin{equation}}
\def\eeq{\end{equation}}

\SetInternalRegister\hbadness{8000} 

\newcommand\doingARLO[2][]{%
  \ifx\mmref\undefined #1\else #2\fi
}

\begin{document}

\title[Dust Polarization From Starlight Data]
{Dust Polarization From Starlight Data}


\author{Pablo Fosalba}{
address={Institut d'Astrophysique de Paris, France}
}

\author{Alex Lazarian}{
address={Department of Astronomy, University of Wisconsin, Madison, USA}
}

\author{Simon Prunet}{
address={Canadian Institute for Theoretical Astrophysics, Toronto, Canada}
}

\author{Jan A. Tauber}{
address={Astrophysics Division, ESA-ESTEC, Noordwijk, The Netherlands}
}

\copyrightyear {2001}

\begin{abstract}

We present a statistical analysis 
of the interstellar medium (ISM) polarization 
from the largest compilation 
available of starlight data, which comprises $\sim 5500$ stars. 
The measured correlation between the mean 
polarization degree and extinction 
indicates that ISM dust grains
are not fully aligned with the uniform component of the 
large-scale Galactic magnetic field.
Moreover, we estimate the ratio of the uniform to the random plane-of-the-sky
components of the magnetic field to be ${\bf B_u/B_r} \approx 0.8$. 
From the analysis of starlight polarization degree and position angle we
find that the magnetic field broadly follows Galactic 
structures on large-scales. 
On the other hand, the angular power spectrum $C_{\ell}$ 
of the polarization degree
for Galactic plane data is found to be
consistent with a power-law, $C_{\ell} \propto \ell^{-1.5}$ 
(where $\ell \approx 180^{\circ}/\theta$ is the multipole order),
for angular scales $\theta > 10^{\prime}$. We argue that this data set
can be used to estimate diffuse polarized emission at microwave frequencies.

\end{abstract}

\date{\today}

\maketitle

\section{Introduction}

The Milky Way Galaxy emits polarized radiation 
at radio, mm-wave, far-infrared and optical wavelengths
(see e.g, \cite{OliTeg99} for a recent review).
The different mechanisms which cause the emission to be polarized
at each of these wavelengths are all related to the Galactic magnetic field.
Therefore the measurement of the polarized
Galactic emission should yield valuable information on our Galaxy's magnetic
field (see e.g, \cite{ZweHei97, Hiletal00, Heietal01}).

Observed starlight polarization is
believed to be caused by selective absorption by 
magnetically aligned interstellar dust grains along the line of sight. 
Since these measurements are limited by dust extinction, they
provide us with a picture of the magnetic field only in the 
vicinity of the sun.
Despite this limitation, recent analyses of such measurements 
(see \cite{Hei00} and references therein) 
suggest that they do contain information about the uniform and 
random components of the magnetic field on large scales.
In particular,starlight polarization vectors trace the plane-of-the-sky
projection of the Galactic magnetic field \cite{ZweHei97} 
and measurements of polarization for stars of different distances reveals
the 3D distribution of magnetic field orientations averaged 
along the line of sight.

The Milky Way magnetic field has also been studied from measurements 
at far-infrared and mm wavelengths \cite{Hiletal99, Novetal00}; see also \cite{Hiletal00, Heietal01}  for recent reviews. 
However, the regions surveyed correspond to few very small 
regions, largely dense dark clouds, mostly in the Galactic plane and
they they reflect rather local distortions of the large-scale magnetic
field, while a global view has only been obtained for external
spiral galaxies similar to our own \cite{ZweHei97}.

At optical wavelengths many polarization measurements 
do exist, which offer an alternative view to our Galaxy.
We shall present below
the most complete compilation to date 
of starlight polarization observations. This analysis will allow us to 
extract basic information on the
large scale statistical properties of the polarization field in the visible.
We do this by studying the correlations between stellar parameters and
computing the angular power spectrum 
of the optical polarization degree from Milky Way stars. 
A more detailed discussion of the analysis and results presented here is
given in \cite{Fosetal01}.

\section{Data}

The starlight polarization data
used in this analysis is taken from the compilation by Heiles 
(see \cite{Hei00} for details and references to the original catalogues; 
see also \cite{Fosetal01}).
This compilation includes data from 9286 sources taken from
a dozen of catalogs
combining multiple observations, providing accurate
positions and reliable estimates for extinction and distance of stars.
From this catalog, we have selected a subsample of 
5513 stars ($60 \%$ of the data) based on the following criteria:
(1) the degree and angle of polarization are given, 
(2) small absolute error in the polarization degree ($< 0.25 \%$)
and (3) a (positive) extinction is given.
All the stars in the Heiles compilation fulfilling the above requirements
also have quoted distance with an estimated 20 $\%$ 
error for most of the sources \cite{Hei00}.
\begin{figure}[t!]	
\includegraphics[angle=90,width=0.6\hsize,height=0.6\hsize]{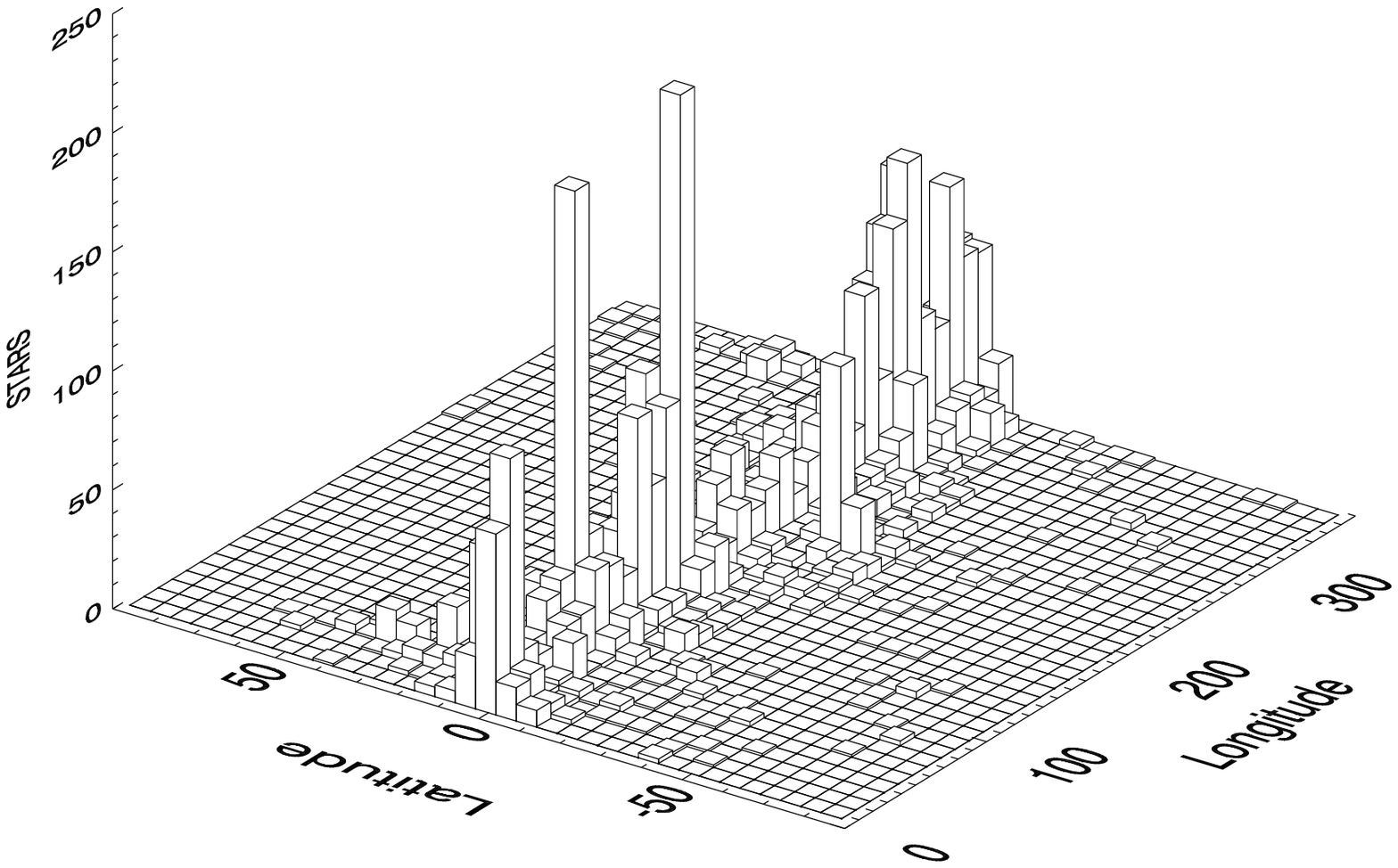}
\hspace{-2cm}
\includegraphics[angle=90,width=0.6\hsize,height=0.6\hsize]{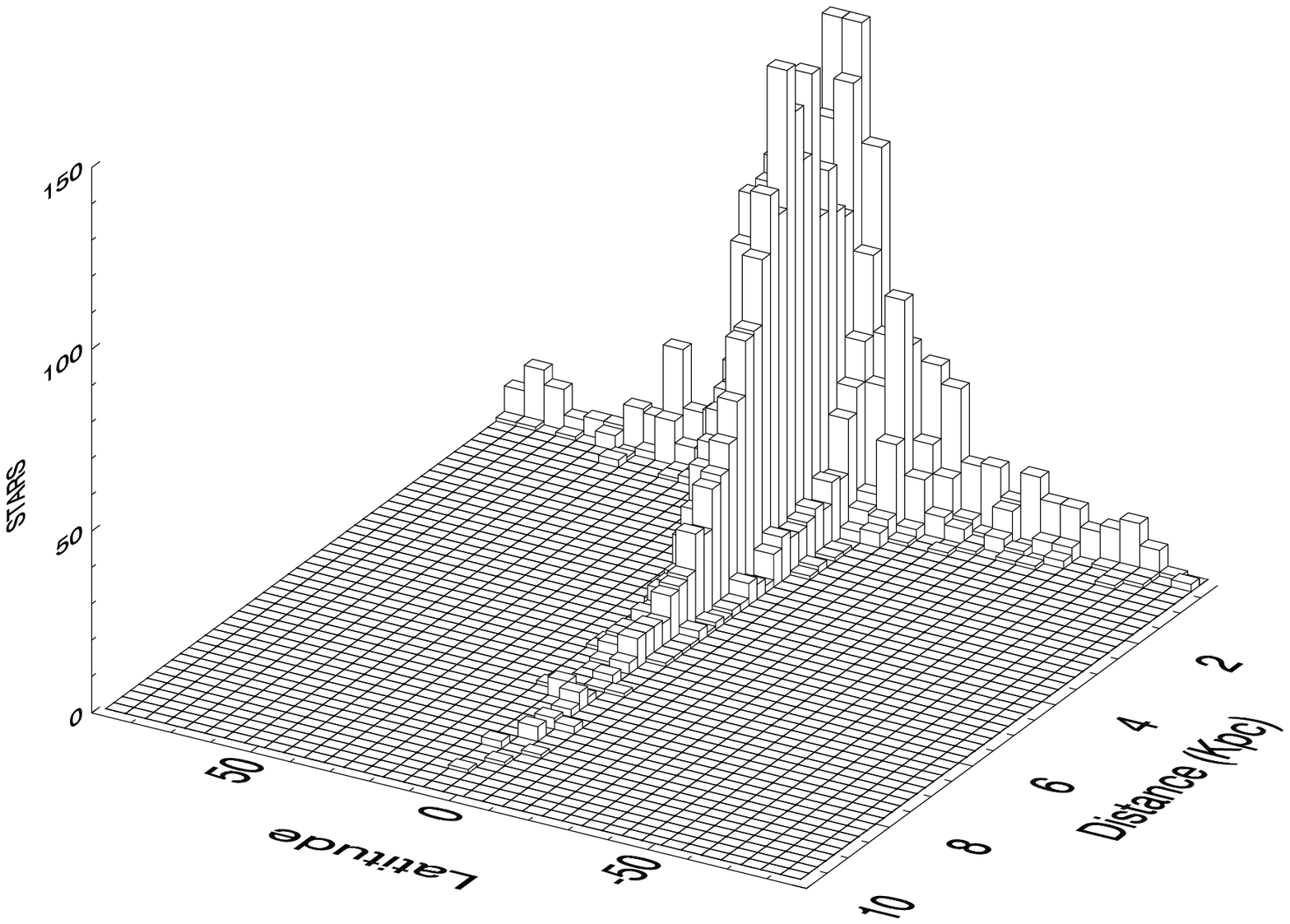}
\caption{\label{hist_star_lb_db}
Distribution of starlight polarization data in Galactic
longitude and latitude bins (\emph{Left}) and distance and latitude bins
(\emph{Right}) for the subsample of 5513 stars analyzed.}
\end{figure}

\begin{center}
\begin{tabular}{ccccc}
\hline
{Latitude} & {Distance} & {Stars $(\%)$} & {P($\%$)} & {E(B-V)}\\ 
\hline
      			 & Total & $4114 (75)$ & $1.69$  & $0.49$ \\
Low Latitude		 & Nearby & $1451 (26)$ & $0.94$ & $0.29$ \\
      			& Distant & $2663 (48)$ & $2.09$ & $0.60$ \\
\hline
  			& Total  & $1399 (25)$ & $0.45$ & $0.15$ \\
High Latitude	        & Nearby & $1315 (24)$ & $0.42$ & $0.14$ \\
              		& Distant & $84 (1)$ & $0.89$ & $0.26$ \\
\hline
\end{tabular}
\end{center}
%\vspace{-0.5cm}
\begin{center}
{\bf Table 1.} Mean Stellar Parameters. High latitude (low latitude) means 
$|{\rm b}| > 10^{\circ}$ ($|{\rm b}| < 10^{\circ}$)
and nearby (distant) denotes d $<$ 1 \kpc ~(d $>$ 1 \kpc).
The quantities between brackets denote amount $\%$ of all stars in the sample.
\end{center}

\begin{figure}[t!]
  \includegraphics[angle=90, width=1.\hsize, height=0.7\hsize]{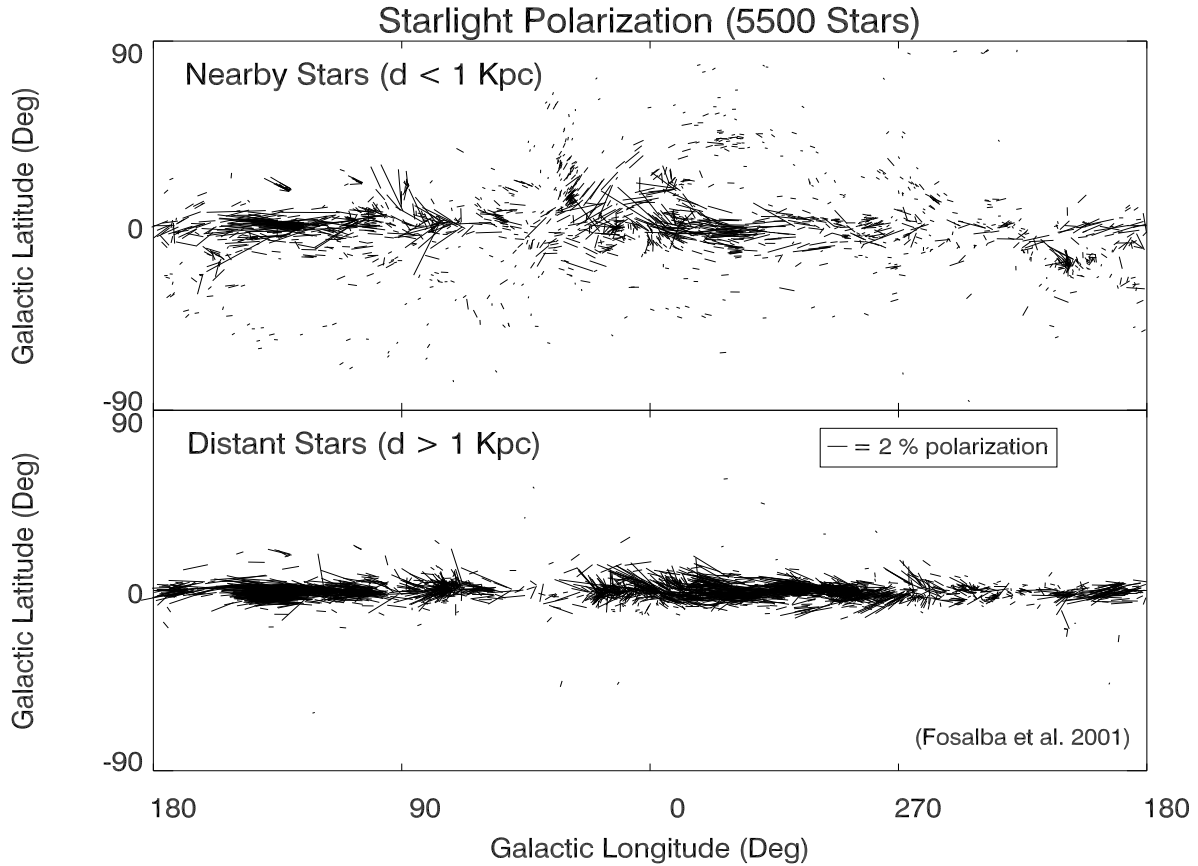}
  \caption{\label{star_polmap}
Starlight polarization vectors in Galactic coordinates for
a sample of 5513 stars. The upper panel shows 
polarization vectors in local clouds, while the lower panel displays 
polarization averaged over many clouds in the Galactic plane.
The length of the vectors is proportional to the polarization degree
and the scale used is shown in the lower panel.}
\end{figure}

Fig \ref{hist_star_lb_db} shows the distribution of sources in our subsample
for data binned in Galactic coordinates (left panel) as well as
in distance and latitude (right panel).
As shown in the latter, practically
all high latitude ($|{\rm b}| > 10^{\circ}$) stars are nearby (d $<$ 1 \kpc).
Within the Galactic plane one can find relatively distant stars, though the
vast majority are within 2 \kpc. 
Therefore, this is a rather local sample.
This is also clearly displayed in the starlight polarization map 
of the subsample of 5513 stars analyzed (see Fig \ref{star_polmap}).

A more quantitative account of this fact is summarized in 
Table 1, where we give the mean stellar parameters 
(i.e, polarization degree P($\%$) and extinction as measured 
by the color excess E(B-V)) in the subsample as a function of 
latitude and distance. It is seen that low-latitude 
stars have large values of the polarization degree P($\%$) $\approx 1.7$, 
and extinction E(B-V) $\approx 0.5$, while high-latitude sources 
exhibit significantly lower values, 
P($\%$) $\approx 0.5$, 
E(B-V) $\approx 0.15$.  
Polarization vectors (defined with respect to Galactic coordinates) 
are typically oriented along the Galactic plane 
($\theta_p \approx 90^{\circ}$) although a more detailed analysis
reveals a rich spatial distribution (see Fig \ref{star_polmap} and 
discussion below).

\section{Starlight polarization and Galactic Magnetic Field}

It is generally accepted that 
grains in diffuse interstellar gas tend to be
aligned with their major axes perpendicular to the magnetic field.
\footnote{
There is no consensus as yet
in relation to what alignment mechanism
is the dominant in the interstellar environment \cite{Laz00}, \cite{LazPru01}. 
Currently, the radiative torque mechanism seems the 
most promising \cite{DraWei96, DraWei97}.} 
According to this picture, the electric field of radiation transmitted 
by an interstellar dust grain is less absorbed along
the grain minor axis and therefore polarized in that direction which is
parallel to the external magnetic field orientation.
Thus, polarized starlight radiation vectors are oriented parallel
to the Galactic magnetic field \cite{ZweHei97}.
This polarization mechanism is usually referred to as 
\emph{differential absorption}. 

Since starlight polarization vectors are only seen as projected 
in the plane of the sky, they just give us direct
information on the plane-of-the-sky projection of the 
Galactic magnetic field orientation.
As shown in  Fig \ref{star_polmap}, there is 
a strong net alignment of starlight polarization vectors 
averaged over many clouds with the Galactic plane
structures (see lower panel) 
as well as a clear 
alignment with the spherical shell of Loop 1 as seen from the 
polarization vectors in local clouds (see upper panel).
This is in remarkable 
qualitative agreement with previous studies, namely \cite{Whi92} 
(based on the catalog of \cite{AxEl76}) and \cite{ZweHei97}. 
In fact, for an homogeneous 
distribution of intervening dust, the larger
the path-length starlight travels to reach the observer, 
the larger the polarization degree
and extinction are expected to be. 
According to this simple picture, regions with measured
low starlight polarization degree (and extinction) correspond to the
\emph{local} ISM,  
while highly polarized ISM regions are observed in the
line-of-sight to distant stars.
This is actually observed in the sample of starlight data we have
analyzed (see Fig \ref{star_polmap}). 
Most of the nearby stars (d $<$ 1 \kpc) 
are found at high Galactic latitudes, while
distant sources (d $>$ 1 \kpc) lie mainly in the Galactic plane.

On the other hand, 
the spatial distribution of the polarization degree and position angle
are expected to be highly correlated and this is also observed in
the starlight polarization map (Fig \ref{star_polmap}), 
where highly polarized regions 
exhibit position angles aligned with the Galactic plane.
There is a (sinusoidal) 
\emph{modulation} of this correlation with Galactic longitude  
due to a projection effect:
one observes the plane-of-the-sky projection of
the polarization vectors that are
aligned with the various Galactic spiral arms 
(see right panel in Fig \ref{star_correl}).

In summary, we find evidence that {\em there is a net alignment of the 
magnetic field (as seen from its plane-of-the-sky projection) 
with Galactic structures on large-scales}.
However, we stress that the full reconstruction of the 
3D magnetic field orientations (and strength) 
requires additional complementary data from radio (synchrotron), 
sub-mm/IR (dust) observations and rotational measures from distant pulsars 
(see \cite{ZweHei97, Han01a, Han01b} and references therein).

\section{Polarization Degree and Efficiency of Magnetic Field Alignment}

\begin{figure}[t!]
 \includegraphics[angle=90, width=0.5\hsize, height=0.5\hsize]{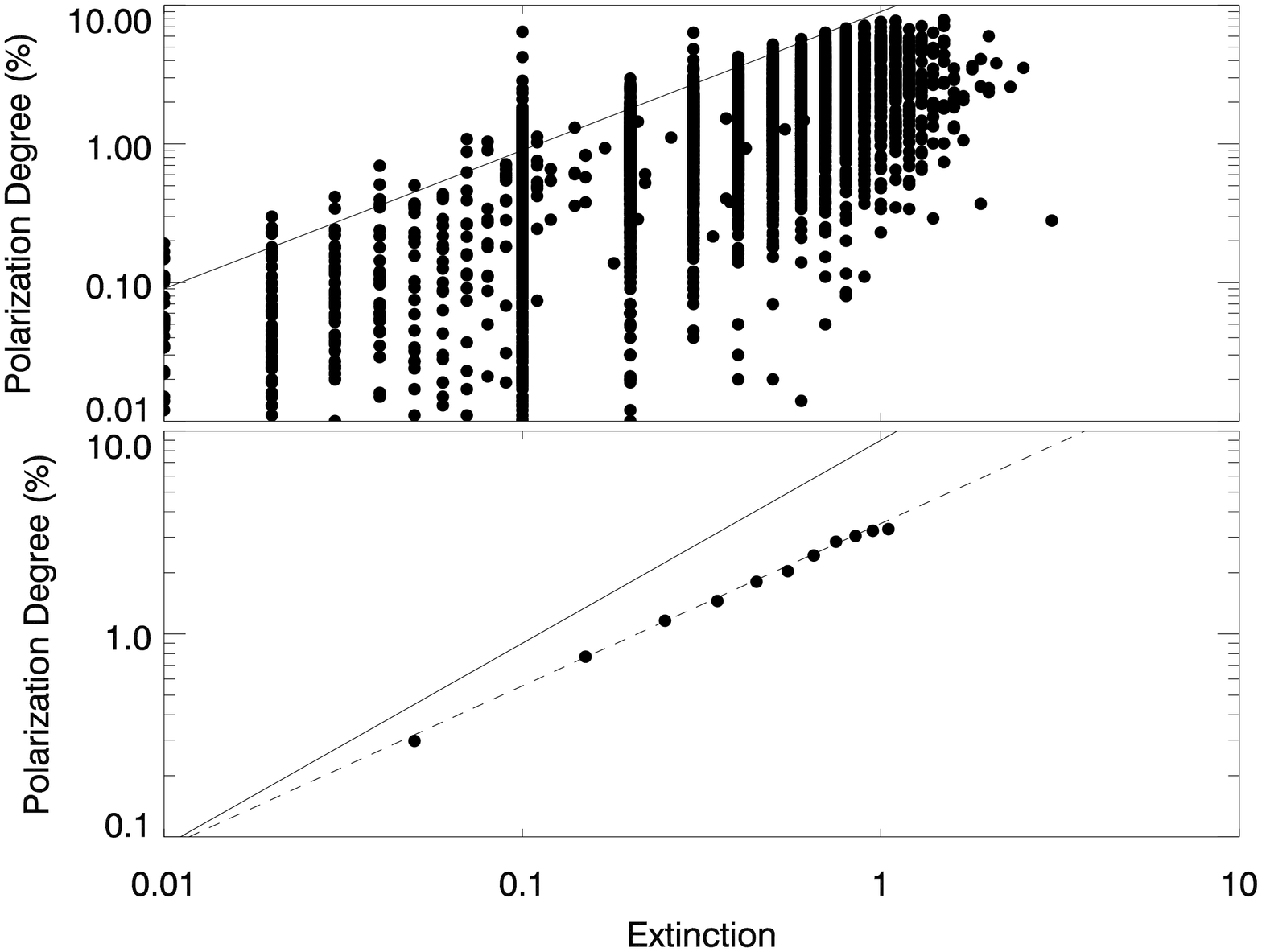}
 \includegraphics[angle=90, width=0.5\hsize, height=0.5\hsize]{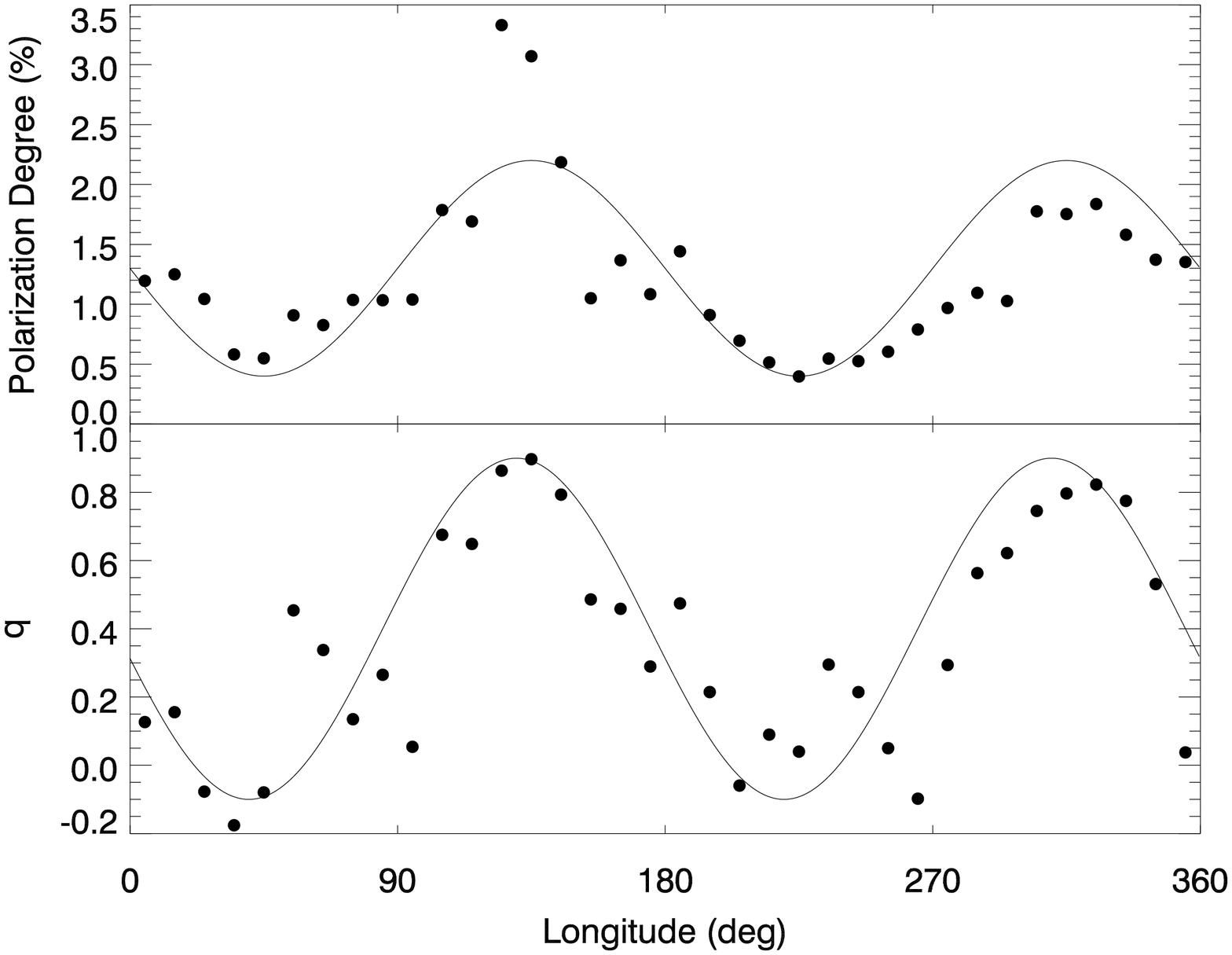}
\caption{\label{star_correl}
(\emph{Left}) 
Correlation between polarization degree P($\%$), and extinction E(B-V).
Upper panel shows all individual sources
while lower panel displays data averaged in extinction bins.
Solid line shows the theoretical upper limit, P($\%$) = 9 E(B-V), for
completely aligned grains by external (regular) magnetic fields.
Dashed line in lower panel shows P($\%$) = 0.39 E(B-V)$^{0.8}$, 
which is a good fit to the data up to E(B-V) $\approx 1$.  
(\emph{Right}) Starlight Polarization Degree (top panel) and
the parameter $q = \cos 2 (\theta_p - 90^{\circ})$ 
(bottom panel) for data averaged in 10$^{\circ}$ 
longitude bins.The solid line shows a best fit to a sinusoidal dependence.} 
\end{figure}

As discussed in the previous section, 
it follows from simple arguments that the 
starlight polarization degree and extinction should be correlated.
We do find such correlation for individual sources in our sample.
However, \emph{on the mean}, the measured correlation,
P($\%$) = 0.39 E(B-V)$^{0.8}$ (see top panel in 
Fig \ref{star_correl}) 
has a lower amplitude than 
what is expected from complete
dust-grain alignment from homogeneous magnetic fields \cite{Jon89}, 
P($\%$) = 9 E(B-V)
(see lower panel in Fig \ref{star_correl}).
The observed roughly linear correlation for individual sources 
is in agreement with measurements at 2.2 $\mu$m \cite{Jon89}
\& 100 $\mu$m \cite{Hiletal95}.
The fact that starlight data exhibits a lower polarization degree 
as a function of extinction
than the theoretical upper limit, suggests that 
{\em either the grain alignment 
is not optimal or the Galactic magnetic field 
has a significant random component}.

In general, we can decompose the total magnetic field as the addition
of a \emph{uniform} (coherent),  ${\bf {B_u}}$, and a \emph{random}
(incoherent) ${\bf {B_r}}$ component.    
A random component of the magnetic field 
smears to some degree the correlation introduced by 
the uniform component \cite{Jon89}. 
This smearing effect is likely to affect the observed stars
as supported by the high degree of incoherence 
observed for the starlight position angle. 
This is especially evident from nearby sources 
(top panel in Fig \ref{star_polmap}).
Assuming that the depolarization is mainly caused by 
the randomness of the magnetic field, one
can relate the observed polarization degree to the ratio of uniform to
random plane-of-the-sky components of the 
underlying magnetic field (see e.g, \cite{Hei96}).
In particular, assuming Burn's model \cite{Bur66}, 
one finds for the starlight sample,
${\bf {B_u}}/{\bf {B_r}} \approx 0.80$, for E(B-V) $\approx 1$, 
corresponding to distant stars 
as an unbiased estimate of the ratio.  
This value is roughly consistent with previous estimates from starlight data 
(see \cite{Hei96} for a review and references therein):
${\bf {B_u}}/{\bf {B_r}} \approx 0.68$, and 
is typically larger than estimates from synchrotron polarization 
or rotational measures of distant pulsars.
The discrepancy 
can be explained as every data set basically samples a different component
of the interstellar medium: 
pulsars mainly trace the warm ionized medium,
starlight data samples primarily the neutral media
while synchrotron data seems to sample all components \cite{Hei96}.

\section{Large-scale Pattern of the Polarized ISM: 
A power spectrum analysis}

The large-scale statistical properties of the
ISM polarization from \emph{absorption} of starlight by dust grains might give
direct statistical information on the polarized diffuse \emph{emission} by
dust: if the grains that extinct starlight and emit constitute the same grain
population, the power spectrum 
of starlight polarization degree is directly related to the power
spectrum of polarized emission from dust. 
Starlight polarization is caused by
aligned grains with sizes $10^{-4}>a>10^{-5}$~cm \cite{KiMa95},
which generate polarized emission in 
diffuse media \cite{PruLaz99}, \cite{LazPru01}. 
Therefore if aligned grains
in diffuse medium have the same temperature the power spectrum 
of the starlight
polarization should be identical to the spectrum of the polarized
continuum from dust in the FIR range (e.g, $100 \mu$m).

In a fully-sampled map, the two-point correlation function
$\xi(\theta)$ of the scalar field ${\rm S}$ 
is simply related to the angular power spectrum (APS):
\beq
\xi (\theta) = <S({\bf q_1}) S({\bf q_2})> = 
\sum_{\ell} {\ell+1/2 \over {\ell(\ell +1)}} C_{\ell} P_{\ell} (\cos \theta)
\label{eq:xi} 
\eeq
where the APS, $C_{\ell}$, 
estimates autocorrelations of the field at an angular scale 
$\theta \approx 180^{\circ}/ \ell$,  
$\ell$ being the so-called multipole order.

We have focused on Galactic plane data ($|{\rm b}| < 10^{\circ}$)
as it concentrates most of the sources in the sample and therefore makes 
the statistical analysis more reliable.
To compute the APS, we use a \emph{hybrid} approach, that we shall
call the \emph{improved correlation function analysis}, that combines
the advantages of real (or pixel) and harmonic (or multipole) space
approaches \cite{SzaSza98, Szaetal01}. 
For this purpose we first compute the two-point correlation function, 
$\xi(\theta)$, of the
polarization degree data using  
a quadratic estimator where shot-noise and edge effects intrinsic 
to the starlight data set are
adequately corrected in pixel space.
\footnote{This method uses the \emph{anafast} program of the HEALpix package
\cite{Goretal98} for the fast computation of the APS. \\
See {\tt http://www.eso.org/science/healpix/}}

\begin{figure}[t!]	
\hbox{
\includegraphics[width=0.5\hsize,height=0.5\hsize]{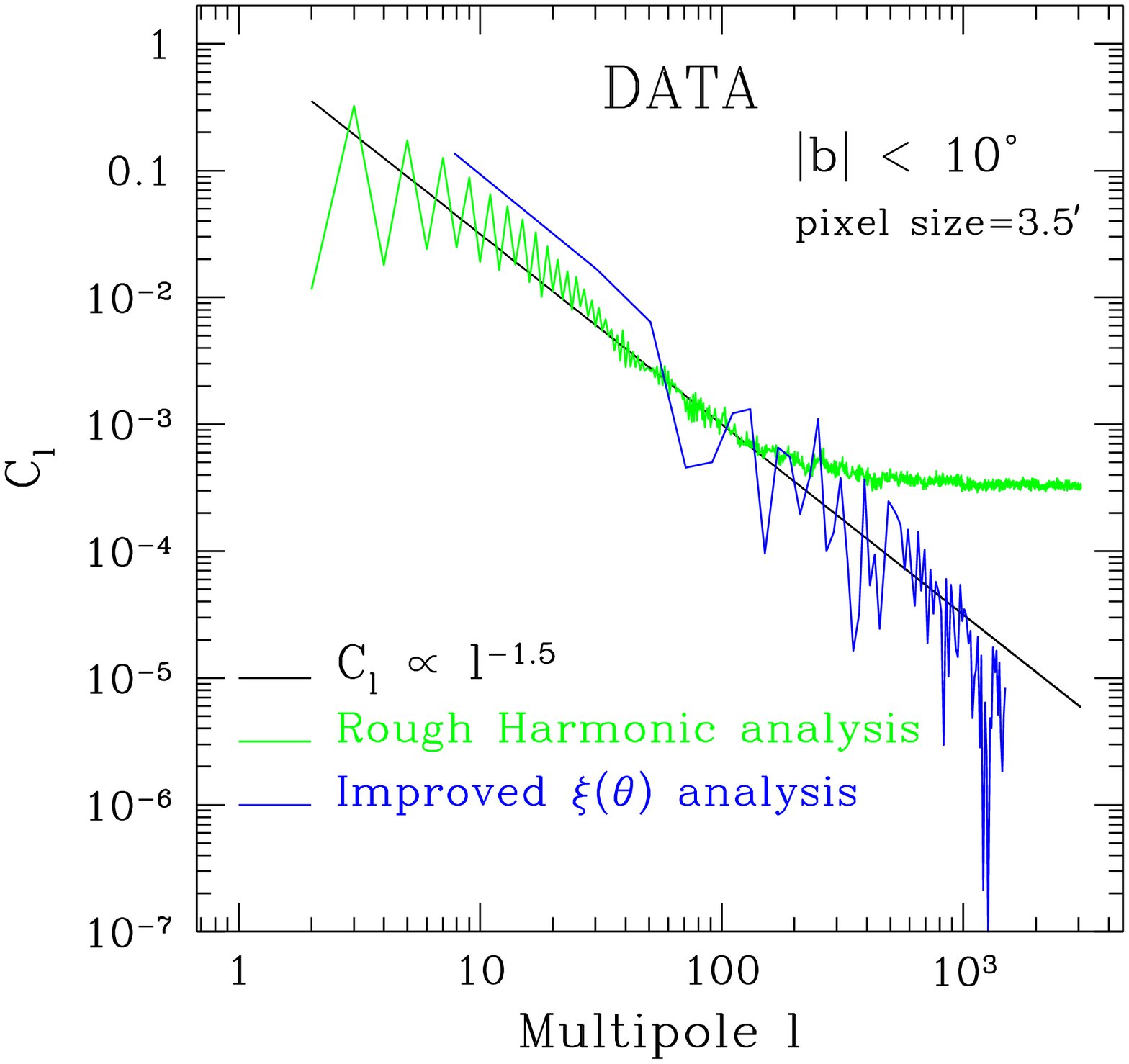}
\includegraphics[width=0.5\hsize,height=0.5\hsize]{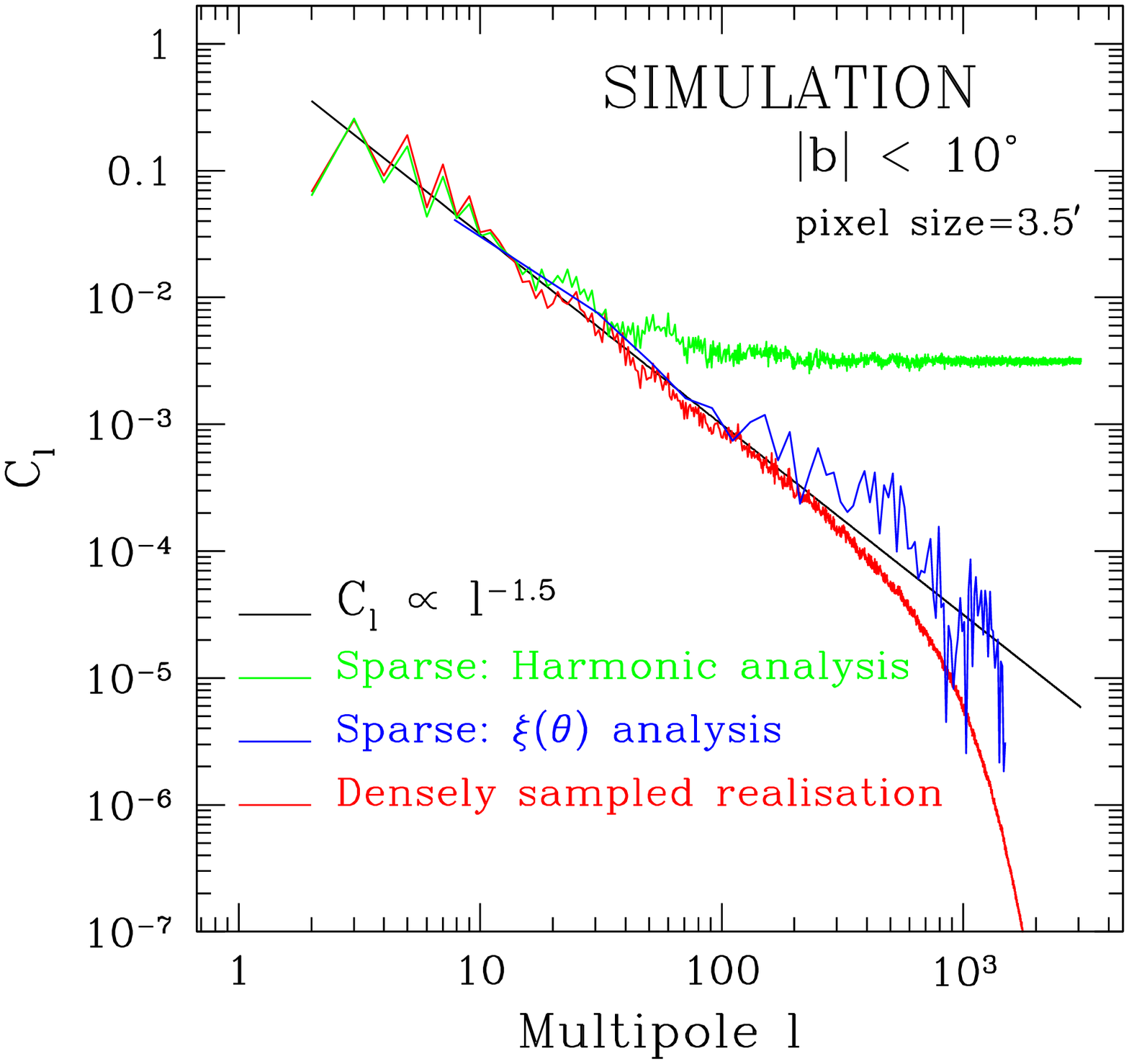}}
\caption{\label{cl_star}
Angular power spectrum  of the starlight polarization degree map
in the Galactic plane, $|{\rm b}| < 10^{\circ}$, for the real data
(\emph{Left}) and the simulated mock catalog (\emph{Right}).
\emph{Rough harmonic analysis} denotes a direct approach in harmonic
space, while \emph{improved $\xi(\theta)$ analysis} is a hybrid approach
(uses techniques both in pixel and harmonic space) that 
corrects for shot-noise and edge effects. As a reference, the red line
shows a densely sample realization of $C_{\ell} \propto \ell^{-1.5}$.}
\end{figure}

Our analysis shows that the starlight polarization degree ${\rm S}$  
is well fitted by a power-law behavior,  
$C_{\ell} \propto \ell^{-1.5}$ for $\ell < 1000$ 
(where the multipole order 
${\ell} \approx 180^{\circ}/\theta$) which translates into 
angular scales ${\theta} > 10^{\prime}$ (see left panel in Fig \ref{cl_star}).
This is approximately the pixel resolution scale used, 
$3.5^{\prime}$. 
We have assessed how the above results are affected by the 
{\em clustering} or non-Gaussianity in the distribution of sources by
simulating a {\em mock starlight map} for the polarization degree 
(see right panel in Fig \ref{cl_star}).
We found that the efficiency with which one measures the 
power spectrum of the 
underlying densely-sampled signal is not significantly altered by 
the clustering of the sources, although for the 
actual data, that is non-Gaussian distributed,
shot-noise dominates at smaller scales 
(i.e, in the direct harmonic approach, $C_{\ell} = Constant$ 
for a larger multipole in the actual data than in the simulation; 
see green lines in Fig \ref{cl_star})
and the estimated power spectrum is noisier
than the simulated random-Gaussian case (see blue lines
in Fig \ref{cl_star}).

The above results provide evidence that the use
of the polarization degree of the ISM as 
sparsely-sampled from 
lines-of-sight to several thousand (Galactic-plane) stars 
allows a clean reconstruction of the APS
of an underlying {\em homogeneously sampled} (continuum) 
polarization degree of the ISM.
In particular, we find that the ISM 
polarization degree in the continuum has the same APS slope
than that measured from sparsely-sampled data,  
$C_{\ell} \propto \ell^{-1.5}$.
It is interesting to note that this slope 
is consistent with that estimated from surveys of 
polarized Galactic synchrotron emission \cite{Tucetal00, Bacetal01},
and the possible underlying common cause certainly deserves investigation.  
In a future work \cite{Fosetal02} we shall discuss how to use
the starlight data set to estimate diffuse polarized emission at 
microwave frequencies
by relating polarization by differential absorption 
in the optical with polarized emission in the sub-mm/FIR 
using dust grain alignment models \cite{Hil88, HilDra95}.
An accurate knowledge of such polarized emission is a critical
issue in the process of component separation in 
cosmic microwave background experiments \cite{PruLaz99, LazPru01}.

%\clearpage

\begin{theacknowledgments}
We acknowledge the use of the starlight data compilation by C. Heiles
who has kindly made it publicly available. 
PF is supported by a CMBNET fellowship of the European Comission.
AL would like to acknowledge the NSF grant AST-0125544.
\end{theacknowledgments}

\doingARLO[\bibliographystyle{aipproc}]
          {\ifthenelse{\equal{\AIPcitestyleselect}{num}}
             {\bibliographystyle{arlonum}}
             {\bibliographystyle{arlobib}}
          }
\bibliography{fosalba}

\end{document}